\documentclass[aps,
twocolumn,
prd,
10pt,
showpacs,
showkeys,
nofootinbib,
longbibliography,
superscriptaddress,
notitlepage]{revtex4-1}
\usepackage{dcolumn}
\usepackage{bm}
\usepackage{ifpdf}
\usepackage{hyperref}
\usepackage{bm}
\usepackage{booktabs}
\usepackage{diagbox}
\usepackage{natbib}
\usepackage{amsmath} 
\usepackage{xcolor, color,graphicx, graphics}
\usepackage[spanish,english]{babel}
\usepackage[latin1]{inputenc}
\usepackage[OT1]{fontenc}
\usepackage{latexsym,amssymb,amsmath,amsfonts}
\usepackage{makeidx}
\usepackage{epsfig,subfigure}
\usepackage{natbib}
\usepackage{epstopdf}
\usepackage{mathrsfs}
\hypersetup{colorlinks=true, linkcolor=blue, citecolor=green}
\usepackage{enumerate}
\usepackage{xcolor}
\usepackage{multirow}
 \usepackage{array}
 \usepackage{graphicx}
\usepackage{subcaption}
\usepackage{acronym}
\usepackage{orcidlink}

\definecolor{red}{rgb}{1,0,0}

\def\+{^\dagger}

\def\<{\leftarrow}
\def\>{\rightarrow}

\def\({\left(}
\def\){\right)}

\def\a{\alpha}    \def\e{\ephilon}
\def\m{\mu} \def\n{\nu}    
\def\k{\kappa}

\newcommand{\bi}{\begin{itemize}} 				\newcommand{\ei}{\end{itemize}}
\newcommand{\benu}{\begin{enumerate}} 		\newcommand{\enu}{\end{enumerate}}
\newcommand{\bd}{\begin{dinglist}{0}}     \newcommand{\ed}{\end{dinglist}}
\newcommand{\bfig}{\begin{figure}[htbp]}  \newcommand{\efig}{\end{figure}}
        			
\newcommand{\bc}{\begin{center}} 				  \newcommand{\ec}{\end{center}}
\newcommand{\be}{\begin{equation}} 				\newcommand{\ee}{\end{equation}}
\newcommand{\bsub}{\begin{subequations}}  \newcommand{\esub}{\end{subequations}}
\newcommand{\ben}{\begin{eqnarray}} 			\newcommand{\een}{\end{eqnarray}}
\newcommand{\ba}[1]{\begin{array}{#1}} 		\newcommand{\ea}{\end{array}}
\newcommand{\bea}{\begin{equation}\begin{array}{rcl}}
\newcommand{\eea}{\end{array}\end{equation}}

\newacro{BH}{black hole}
\newacro{ADAF}{advection-dominated accretion flow}
\newacro{PW}{Paczy\'nski-Wiita}
\newacro{GR}{General Relativity}
\newacro{AGN}{active galactic nucleu}
\newacro{XRB}{X-ray binary}
\newacro{GRB}{gamma-ray burst}
\newacro{EHT}{Event Horizon Telescope}
\newacro{TDE}{tidal disruption event}
\newacro{RBGs}{Ricci-Based gravity theories}

\def\k{\kappa}   \def\m{\mu} \def \e{\epsilon} \def\a{\alpha} \def\n{\nu}

\begin{document}


\title{Effects of Eddington-inspired Born--Infeld Gravity on Black Hole Accretion}
\author{Merce Guerrero \orcidlink{0000-0001-9888-6943}}
\affiliation{Institute of Physics, University of Tartu, Estonia}
\author{Razieh Ranjbar \orcidlink{0000-0003-1488-4890}}
\affiliation{School of Astronomy, Institute for Research in Fundamental Sciences (IPM), P.O. Box 19395-5531, Tehran, Iran}


\begin{abstract}
 Black holes offer a unique setting to test gravity in the strong-field regime, where deviations from General Relativity are expected to be most pronounced. We study how modified gravity changes the structure of an advection-dominated accretion flow (ADAF) around a supermassive black hole. The gravitational field governing the disc is described by a Paczy\'{n}ski--Wiita-like potential derived in Eddington-inspired Born--Infeld (EiBI) gravity, formulated \textit{\`a la} Palatini. We solve the ADAF-like equations for this potential and compare the radial profiles of the flow with those obtained for the standard Paczy\'{n}ski--Wiita potential. The largest difference appears in the gas density, which is several orders of magnitude higher in the EiBI case near the inner region and converges to the standard result at large radii. The specific angular momentum and temperature are lower in EiBI gravity, while the radial velocity differs mainly at large radii for both solution branches, $ s=-1$ and $+1$. The angular velocity and the disk aspect ratio $H/r$ change only slightly. These results show that EiBI gravity can modify the density structure of hot accretion flows in the strong-field regime, potentially providing observational discriminators.
\end{abstract}

\maketitle

\section{Introduction}

Black holes (BHs) are among the most remarkable and intriguing predictions of Einstein's theory of \ac{GR}. The event horizon of a \ac{BH} is extremely difficult to observe directly in any wavelength, since it does not radiate. However, when it is surrounded by material, it provides the central gravitational environment that leads to a wide variety of accretion phenomena \citep{AbramowiczFragile2013}, including \acp{AGN}, \acp{TDE}, and \acp{XRB} \citep{AbramowiczFragile2013}. 

The accreting material radiates a large fraction of its energy near the black hole, and this emission spans the electromagnetic spectrum. Hot gas in the inner flow produces X-rays \cite{Remillard2006}, the accretion disk emits mainly in the optical and ultraviolet in AGN \cite{Koratkar1999,Padovani2017}, and jets and radiatively inefficient flows contribute synchrotron emission from radio to infrared \cite{Yuan2014}, with high-energy emission extending to gamma rays in some sources \cite{Blandford2019}. Black holes are therefore observed indirectly through this multiwavelength emission, in systems such as X-ray binaries, active galactic nuclei, and relativistic jets \cite{Frank2002}.

To describe the dynamics of matter in these environments, one must, in principle, account for the strong gravitational field near the \ac{BH} using \ac{GR}. However, fully relativistic treatments are often computationally demanding. For this reason, simplified approaches have been developed to model accretion disks that incorporate key relativistic effects into a Newtonian framework \cite{Kato2008}.

Inside these semi-analytic models, the structure of accretion flows around compact objects depends on the temperature, which at the same time divides them into two classes. In cold accretion flows, as the accretion rate increases, radiative cooling becomes more efficient, and the flow can transition toward an optically thick, geometrically thin disk (\cite{ShakuraSunyaev1973}). At accretion rates approaching or exceeding the Eddington rate, photon trapping and radiation pressure become increasingly important, producing a geometrically thicker slim-disk configuration (\cite{Abramowicz1988}. Conversely, hot accretion flows, at low accretion rates, the flow is typically hot, optically thin, and geometrically thick, with inefficient radiative cooling and substantial energy advection (\cite{Ichimaru1977}; \cite{NarayanYi1994, NarayanYi1995}). The first hot accretion flow model was proposed by (\cite{ShapiroLightmanEardley1976}; hereafter SLE), characterised by optically thin gas with temperatures approaching the virial value, but it was subsequently found to be thermally unstable \cite{Ichimaru1977}. This emphasised the importance of advection, showing that viscously dissipated energy can heat the flow rather than being radiated away, and argued that the resulting hot accretion solution is thermally stable.

 In hot accretion flows, most of the gravitational energy released during accretion is advected into the \ac{BH} rather than radiated away, leading to the well-known \ac{ADAF} model \cite{Yuan2014}. However, recent \ac{EHT} observations \cite{EHT2021} and numerical simulations suggest that the mass accretion rates onto \acp{BH} are lower than those predicted by the classical \ac{ADAF} model, mainly due to mass loss through winds and outflows \cite{StonePringleBegelman1999, Narayan2012, YuanBuWu2012, LiOstrikerSunyaev2013}. Motivated by these results, we adopt an \ac{ADAF}-like accretion flow with low angular momentum \cite{Narayan2011, Ranjbar2022, Ranjbar2025}. This model provides an intermediate solution between Bondi and \ac{ADAF} accretion and predicts lower mass accretion rates \cite{Narayan2011, Park2009, HanPark2026, Jiao2026}, in better agreement with observations. Such models apply to various high-energy astrophysical systems, including \acp{TDE}, \acp{GRB}, and accretion flows around supermassive \acp{BH} such as M87* and Sgr A*. 

Despite their simplicity, semi-analytic models account for relativistic effects, for example, in describing the global dynamics and energetics of a wide variety of accretion systems. A prominent example is the standard thin-disk model of \cite{ShakuraSunyaev1973}, formulated within a Newtonian framework, with relativistic effects often incorporated by adopting the marginally stable circular orbit as the disk's inner termination radius. Similarly, the thick-disk model of \cite{Paczynski1980} retains the Newtonian dynamical formulation while replacing the classical gravitational potential, $\Phi_{\rm N}=-GM/r$, with a pseudo-Newtonian potential designed to reproduce key properties of the relativistic gravitational field in the strong-gravity regime near a \ac{BH}. The approach adopted here is motivated by the relativistic treatments of perfect-fluid disks around \acp{BH} developed by \cite{Abramowicz1978} and \cite{Kozlowski1978}, with the earlier work of \cite{FishboneMoncrief1976} providing an important foundation for this class of models. 

One of the most widely used pseudo- Newtonian potentials for non-rotating \acp{BH} is the \ac{PW} \cite{Paczynski1980}. It successfully reproduces several essential features of the Schwarzschild spacetime, including important aspects of orbital motion in the strong-field region, while maintaining the simplicity of Newtonian calculations. Although it does not capture all relativistic effects exactly, it provides a practical and efficient approximation that is especially useful in studies of hot accretion flows, where a fully general-relativistic treatment is not always required. Several alternative approaches have subsequently been developed to incorporate relativistic effects while retaining the simplicity of a Newtonian framework, including modified Newtonian potentials and generalised Newtonian formulations for accretion flows around \acp{BH} \cite{Ranjbar2019, Tejeda:2013mva}. These approaches, however, remain phenomenological: the potential is adjusted to reproduce relativistic orbital properties without being derived from an underlying modification of the gravitational field equations themselves.

These pseudo-Newtonian potentials share a common underlying assumption, namely that the background spacetime is that of standard \ac{GR}. An alternative strategy is to relax this assumption 
and to construct the effective potential from the resulting solution of modified gravitational theories. Such modifications can be particularly relevant in high-curvature or high-density environments, where deviations from \ac{GR} predictions may be appreciable. Consequently, compact objects and their surrounding accretion flows provide promising astrophysical laboratories for investigating the phenomenological consequences of modified gravity and potentially constrain them.

Among the various approaches to extending \ac{GR}, we will focus on a particular theory within a class of theories known as \ac{RBGs}. These theories are formulated within the Palatini approach, in which the metric and affine connection are treated as independent variables. Additionally, their Lagrangians are constructed from scalar invariants obtained through contractions of the metric with the torsion-free Ricci tensor. As a result, their field equations remain second-order and do not introduce additional propagating degrees of freedom. These theories are consistent with current Solar System and gravitational-wave observations while also allowing for possible deviations from GR in the strong-field regime. In this way, \ac{RBGs} allow for nontrivial modifications of gravitational dynamics while preserving a relatively simple geometric description of gravity. This simplicity has enabled numerous studies of different types of stars \cite{Pani:2011mg, Delsate:2012ky, Casanellas:2011kf, Avelino:2012ge, Banerjee:2017uwz, Wibisono:2017dkt, Danarianto:2019mxf, Rosyadi:2019hdb, Banerjee:2021auy, Guerrero:2021fnz}, thereby providing a significant framework for testing the effects on accretion disks.

In this work, we investigate the effects of EiBI gravity using the exact solution found in \cite{Guerrero2020EiBIRotating}, which introduces modifications to \ac{GR}, on low-angular-momentum \ac{ADAF}-like accretion flows. We explore how deviations from \ac{GR} in the gravitational potential affect the properties of these accretion flows and their observational implications. Throughout this work, we adopt units in which $r_{\rm s} = c = \rho_{\rm out} = 1$.

The paper is structured as follows: Section \ref{sec:2} presents the general framework of EiBI gravity, an exact solution, and the derivation of the \ac{PW}; Section \ref{sec:3} describes the structure of the ADAF-like accretion flow and the numerical method used to obtain the solutions. Section \ref{sec:4} presents our numerical results. In contrast, Section \ref{sec:5} provides a discussion of the main results and conclusions.

\section{General framework}\label{sec:2}

In this section, we introduce the main gravitational ingredients that will be employed to construct the semi-analytic ADAF accretion disk model. We begin with a brief overview of EiBI gravity, outlining its formulation and assumptions. We then discuss an exact spherically symmetric solution obtained using the so-called \textit{mapping} method. Finally, we introduce a pseudo-Newtonian potential derived from this solution, which will serve as the basis for modelling the accretion disk in the following section.

\subsection{Eddington-inspired Born-Infeld gravity}

The action of EiBI gravity is given by
\begin{eqnarray}
\mathcal{S}_{EiBI}&=&\frac{1}{\kappa^2 \epsilon} \int d^4x \left[\sqrt{-\vert g_{\mu\nu} + \epsilon R_{\mu\nu}(\Gamma) \vert} - \lambda\sqrt{ -g}\right] \nonumber \\
&+&  \mathcal{S}_m (g_{\mu \nu}, \psi_m) \label{eq:SEiBI}
\end{eqnarray}
where $\kappa^2=8\pi G$ is Newton's constant, $\e$ is the (length squared) EiBI parameter, $\lambda$ is related to the asymptotic character of the solutions (from now on we fix $\lambda=1$ to deal with asymptotically flat solutions). Vertical bars denote the determinant, but also $g$ denotes the determinant of the spacetime metric $g_{\mu\nu}$. We are using the Palatini formalism, in which the metric is independent of the affine connection $\Gamma \equiv \Gamma_{\mu\nu}^{\lambda}$ from which the (symmetric) Ricci tensor is calculated. 

The field equations can be written as \cite{BeltranJimenez:2017doy}
\begin{equation}\label{Eq: EiBI Ricci tensor}
{R^\mu}_\nu (q) = \dfrac{1}{\vert \hat{\Omega}\vert^{1/2}}\left(  \, \frac{\vert \hat{\Omega} \vert^{1/2}-\lambda}{\epsilon} {\delta^\mu}_\nu+ \k^2{T^\mu}_\nu \right) \ ,
\end{equation}
where
\begin{eqnarray}
    q_{\mu \nu} \equiv g_{\mu\nu} + \epsilon \, R_{\mu\nu}(q) \ , \label{Eq: q 1} \\
      |\Omega|^{1/2} ({\Omega^\m}_\nu)^{-1} = \delta_{\mu}^\nu -\k^2 \e \, {T^\m}_\n \ . \label{Eq: deformation matrix}
\end{eqnarray}
Note that Eq. \eqref{Eq: EiBI Ricci tensor} shows the existence of an Einstein frame for this theory, sourced on its right-hand side via additional couplings in the matter fields. This is a common feature of a more general family of theories called Ricci-Based Gravities (RBG) \cite{Afonso:2018bpv}. Another property of this group of theories is that the Einsteinian metric, $q_{\mu\nu}$, is compatible with the independent connection. Additionally, the {\it deformation matrix}, ${\Omega^\m}_\a$, relates the Einsteinian metric, $q_{\mu\nu}$, with the spacetime metric, $g_{\mu\nu}$, by the algebraic relation

\begin{equation}\label{Eq: q Omega}
q_{\mu \nu}= g_{\mu\a} {\Omega^\a}_\n \ .
\end{equation}

From the above equation, one can see that if we do not couple the theory to a matter field, the deformation matrix \eqref{Eq: deformation matrix} is the identity, which reduces to GR. Therefore, for EiBI gravity and other RBGs, having a matter Lagrangian is necessary to differ from GR.

\subsection{Spherically symmetric black holes}

In Refs. \cite{Guerrero:2021avm, Guerrero2020EiBIRotating, Afonso:2018bpv, Afonso:2018mxn, Afonso:2018hyj, Shao:2020weq}, the authors obtain an exact solution using the so-called mapping method. This approach exploits the structure of the field equations \eqref{Eq: EiBI Ricci tensor} to establish not only a correspondence between the EiBI gravity and GR field equations, but also between their respective solutions. As a result, the problem is reduced from solving a system of highly non-linear differential equations to solving a set of algebraic relations.

As an example, ref. \cite{Guerrero2020EiBIRotating} shows how the Maxwell electrodynamics coupled to GR maps to a Born-Infeld electrodynamics coupled to the EiBI gravity. At the same time, the solutions in both gravitational theories are consistent with the mapping of the solutions: from the Reissner-Nordstr\"om metric in GR used as a seed to obtain the following metric in EiBI gravity,

\begin{eqnarray}
d{s}^2&=&-\frac{A(r)}{\Omega_+}dt^2+\frac{dr^2}{A(r)\Omega_+}+r^2(x)(d\theta^2+\sin^2\theta d\varphi^2) \ , \label{eq:gmn-F}\\
A(r)&=&1-\frac{2GM(r)}{x(r)} \ , \label{eq:A(x)} \\
M(r)&=&M_0 -\frac{ r_q^2}{2x(r)  } \ , \label{eq:Mp}\\
\Omega_\pm&=& \frac{1}{1\pm\frac{s r_c^4}{x^4}}\ , \\ \label{eq:Omegapx}
x^2(r)&=& \frac{1}{2}\left( r^2 +\sqrt{r^4+4sr_c^4}\right) \ . \label{eq:Omegamx}
\end{eqnarray}

Several constants characterise this solution. First, there is the critical radius, $r_c^4\equiv|\epsilon|\kappa^2 Q^2/8\pi$, which corresponds to the wormhole throat when $s=-1$, while for $ s=1$, it determines the value of $x$ at $r=0$. Second, the charge radius is defined as $  r_q^2 = Q^2/(4 \pi \epsilon_0)$. Finally, $\epsilon_0$ and $M_0$ are the electric constant and an integration constant that can be identified with the asymptotic Schwarzschild mass, respectively. Note that setting $ r_c \rightarrow 0$, we recover the Reissner-Nordstr\"om solution of GR.

\subsection{PW-like potential}

In this section, we focus on the motion of test particles. Assuming particles follow spacetime geodesics, their trajectories are completely determined by the metric. Therefore, the PW potential is constructed solely from the metric and does not depend explicitly on other geometric quantities. Although the PW potential has been extended to rotating spacetimes, we restrict our attention to a static, spherically symmetric case, following the original motivation. In such a case, the particle motion reduces to a one-dimensional radial effective potential.

The PW potential reproduces the marginally stable and marginally bound radii, and, with less precision, the specific angular momentum and the binding energy for circular orbits \cite{Tejeda:2013mva}. Consequently, this potential offers a compelling example of a pseudo-Newtonian potential, intended to approximate the actual effective potential. It was first introduced in ref. \cite{Paczynski1980}, however, this concept was not mathematically derived there. In a subsequent work, M. A. Abramowicz showed a ``step-by-step" derivation of the potential that was reused and generalised to any static, spherically symmetric spacetime in \cite{Faraoni:2015lud}\footnote{Note that the choice of pseudo-Newtonian potential is not unique. Actually, there are many other proposals \cite{Tejeda:2013mva, Tejeda:2014oxa,Nowak-Wagoner}}. 

The PW-like potential can be calculated with the following relation \cite{potential, Faraoni:2015lud}
\begin{equation}\label{Eq: PW general}
   \Phi_{PW}=\frac{1}{2} (g^{tt} +1) \ ,
\end{equation}
where we rewrite the inverse of the time component metric in Eq. \eqref{eq:gmn-F} as
\begin{eqnarray}
 g^{tt}(r)=-\frac{\Omega_+}{A(r)}=-\Omega_+\left(\frac{2GM(r)}{x(r)-2GM(r)}+1\right) \ .
\end{eqnarray}

\begin{figure}
    \centering
    \includegraphics[width=0.9\linewidth]{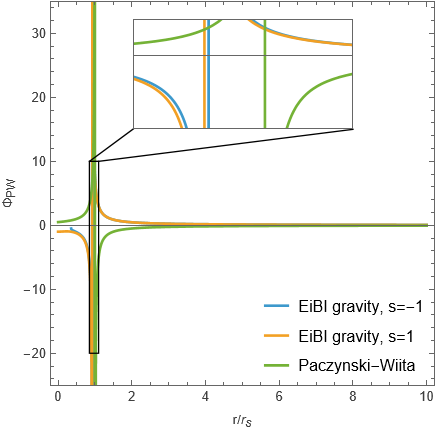}
        \caption{Comparison of the PW-like potential for the two branches of the EiBI gravity solutions and the standard Schwarzschild results. We chose $r_c=0.5$ and $r_q=0.5$, and, together with the radial coordinate $r$, they are normalised by $r_s$. }
    \label{fig:PW}
\end{figure}

The pseudo-Newtonian potential is chosen so that the extrema of the Newtonian effective potential coincide with those for timelike geodesics of the spacetime. In Fig. \ref{fig:PW}, we show the difference between $s=\pm1$, using the same values for the solution parameters, $r_c/r_s$ and $r_q^2/r^2_s$, as compared with the original PW.

An important feature that the PW potential has to have is its limit to the Newtonian potential at large distances, $r\rightarrow\infty$. For this, one can check that $\Omega_+$ tends to 1 and $ M(r)$ to a constant $M_0$ in such a limit. Thus, the Newtonian limit of the PW potential \eqref{Eq: PW general} tends to the correct value,
\begin{equation}
     \Phi_{PW}|_{r\rightarrow\infty}=\frac{GM_0}{r} \ .
\end{equation}

\begin{figure}
        \centering
        \includegraphics[width=1.0\linewidth]{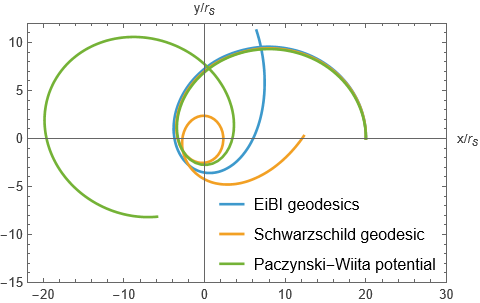}
        \caption{Trajectories of a timelike particle in a Schwarzschild and EiBI spacetime solving the geodesic equation, and under the PW potential written in Eq. \eqref{Eq: PW} using the classical Hamiltonian equation. For EiBi gravity, we use the same parameter values as in Fig.\ref{fig:PW}}
        \label{fig:trajectories}
    \end{figure}

Similarly to Newtonian physics, we can define an effective potential for the orbital motion as \cite{potential}

\begin{equation}\label{Eq: Veff}
    V_{eff}(r,L)\equiv \frac{1}{2} \ln{\left[g^{tt}+L^2 g^{\phi\phi}\right]}.
\end{equation}
Even though we call it an effective potential, it is worth pointing out that this potential is not the same as the full geodesics, since $g_{tt}\neq g^{rr}$. As a consequence, instead of having the usual $\dot{r}^2/2$, one actually has a function multiplying the radial velocity. Despite this, the actual effective potential and the one defined in Eq.~\eqref{Eq: Veff} share the same extrema. Therefore, the location of circular orbits coincides, which corresponds to when $\dfrac{d V_{eff}(r)}{dr}=0$. An example of orbital differences can be observed in Fig. \ref{fig:trajectories}, where we use the same energy and initial velocity for the geodesics of EiBI and Schwarzschild, but a different energy for the PW to obtain a closed orbit.

From the Eq. \eqref{Eq: Veff}, the Keplerian angular momentum, $L_{PW}$, can be obtained giving the result
\begin{eqnarray}
L^2_{PW}(r)=
\end{eqnarray}
\begin{equation*}
    \frac{r^2 \left(r (r-2G M(r)) \Omega'_+ -2 \Omega_+ \left(GM(r)-r GM'(r)\right)\right)}{2 (r-2 GM(r))^2} \ .
\end{equation*}

Finally, the specific angular momentum, $\Omega_{PW}$, is found as usual

\begin{equation}
    \Omega_{PW}= 
\end{equation}
\begin{equation*}
    \frac{\sqrt{ r (r-2 GM(r)) \Omega'_+-2 \Omega_+ \left(GM(r)-r GM'(r)\right)}}{\sqrt{2}  (r-2 GM(r))} \ ,
\end{equation*}

where prime denotes the derivative with respect to the radial coordinate. The expansion of the above quantities at large distances also boil down to the Newtonian result,
\begin{align}
    L^2 =& \frac{ 2 G M_0}{4}\,r +O\left(\frac{1}{r}\right)^0 \ , \\
     \Omega_{PW}= & \frac{\sqrt{ 2 G M_0}}{2} \left(\frac{1}{r}\right)^{3/2} +O\left(\frac{1}{r}\right)^{5/2} \ ,
\end{align}
as was expected.


\section{ADAF-like equations}\label{sec:3}

In this section, we will define all the equations required to describe the \ac{ADAF}-like accretion flow based on \cite{Narayan2011}. 

The general form of the continuity equation is
\begin{equation}
    \frac{\partial \rho}{\partial t} + \text{div} (\rho \boldsymbol{v}) = 0,
\end{equation}
where $\rho$ denotes the mass density and $\boldsymbol{v}$ is the spatial velocity vector. Note that bold letters mean vectors.

Throughout this work, we assume no vertical outflow, steady-state flow ($\partial/\partial t = 0$), and axisymmetry ($\partial/\partial \phi = 0$). Under these assumptions, the continuity equation in cylindrical coordinates reduces to
\begin{equation}\label{Eq: Cont eq}
    \frac{1}{r} \frac{\partial}{\partial r} (r \rho v_r) = 0 \ .
\end{equation}

This implies that the quantity $r \rho v_r$ does not depend on the radius.

Integrating over the azimuthal direction and over the vertical extent of the equation \eqref{Eq: Cont eq}, from $-H$ to $H$, and assuming that $\rho$ and $v_r$ represent vertically averaged quantities, we obtain that the mass accretion rate is
\begin{equation}
\dot{M}
=-\int_0^{2\pi}\int_{-H}^{H}\rho v_r r,dz,d\phi
=-4\pi rH\rho v_r\ .
\end{equation}
Thus, the steady-state continuity equation implies that the mass accretion rate is independent of radius.

The Newtonian hydrodynamic equation of motion is given by
\begin{equation}
    \rho \frac{d\boldsymbol{v}}{dt} 
    = - \boldsymbol{\nabla} P - \rho \boldsymbol{\nabla} \Phi + \rho \boldsymbol{N},
\end{equation}

where $\boldsymbol{N}$ represents the viscous force, $P$ the pressure of the gas and $\Phi$ is the gravitational potential. In addition,
$\rho \boldsymbol{N}$ is the viscous force per unit volume, given by 
\begin{equation}
    \rho N_i = \frac{\partial t_{ik}}{\partial x_k}
\end{equation}

where $t_{ik}$ is the viscous stress tensor.

For the accretion disk, only the $r\phi$-component of the viscous stress
tensor, $t_{r\phi}$, is dominant; therefore, $N_r = N_z = 0$. 

By adopting the new potential, the gravitational potential of the \ac{BH} in Eq. \eqref{Eq: PW general} can be expanded to
\begin{align}\label{Eq: PW}
\mathbf{\Phi}_{\rm PW}(r) = \\
&-\left(1+\frac{s r_c^4}{x(r)^4}\right)
\left(
\frac{r_s x(r)-r_q^2}
{-r_s x(r)+x(r)^2+r_q^2}
+1
\right),\nonumber
\end{align}

where $r_{\text{s}}$ is the Schwarzschild radius of the body given by $r_s = 2GM/c^2$. 

The Keplerian angular velocity in this potential is

\begin{equation}
\begin{split}
\Omega_K
&=
\frac{\sqrt{r^3x(r)}}
{2r^2\left(-r_sx(r)+x(r)^2+r_q^2\right)x(r)^{3/2}}
\\
&\quad\times
\Bigl[
-3r_c^4r_ssx(r)
+4r_c^4sx(r)^2
-2r_q^2x(r)^4
\\
&\qquad\quad
+r_sx(r)^5
+2r_c^4r_q^2s
\Bigr]^{1/2}.
\end{split}
\end{equation}

\begin{equation}
\Omega^2 = \frac{v_{\phi}^2}{r^2}, \quad p = \rho c_s^2.
\end{equation}

Here, $c_s$ denotes the isothermal sound speed. 
The steady-state radial and $\phi$ component of the momentum equation of the accreting gas in the presence of viscosity become, respectively
\begin{equation}\label{eq: radial equation}
    v_r \frac{dv_r}{dr} = (\Omega^2 - \Omega_K^2) r - \frac{1}{\rho} \frac{d}{dr}(\rho c_s^2),
\end{equation}

\begin{equation} \label{eq: angular equation}
    v_r \frac{d}{dr} (r^2 \Omega)  = \frac{1}{\rho r H} \frac{d}{dr} (\nu \rho r^3 H \frac{\partial \Omega}{ \partial r}) \ .
\end{equation}

The kinematic viscosity is parameterised using the $\alpha$ prescription,
\[
\nu = \alpha \frac{c_s^2}{\Omega _K}.
\]

By integrating  $\phi$-component of momentum equation (Eq. \eqref{eq: angular equation}), we obtain angular velocity
\begin{equation}\label{int: angualr eq}
    \frac{d\Omega}{dr} = \frac{v_r\Omega_K (\Omega r^2 - j)}{\alpha c_s^2 r^2} \ ,
\end{equation}

where $j$ the specific angular momentum swallowed by the \ac{BH}. The energy equation for the accreting gas is given by

\begin{equation}\label{energy equation}
    \frac{\rho v_r}{\gamma -1} \frac{d c_s^2}{dr} - c_s^2 v_r \frac{d \rho }{dr} = \frac{f \alpha c_s^2 \rho r^2}{\Omega_K} \left(\frac{d \Omega}{dr}\right)^2.
\end{equation}

where $f$ represents the fraction of viscous heat that is advected with the flow, and $\gamma$ is the adiabatic index. Here, we adopt $f=1$ and $\gamma=5/3$.

\begin{figure*}[htbp]
    \centering

    \includegraphics[width=0.33\linewidth]{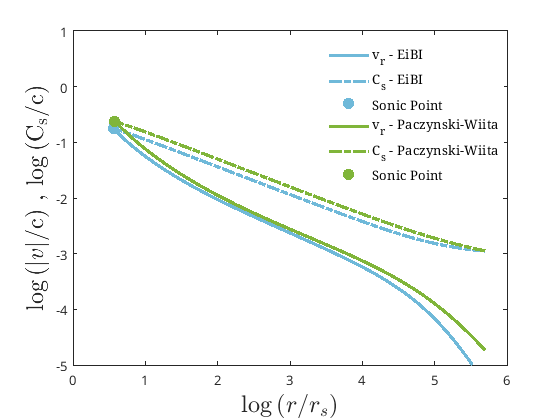}
    \hfill
    \includegraphics[width=0.32\linewidth]{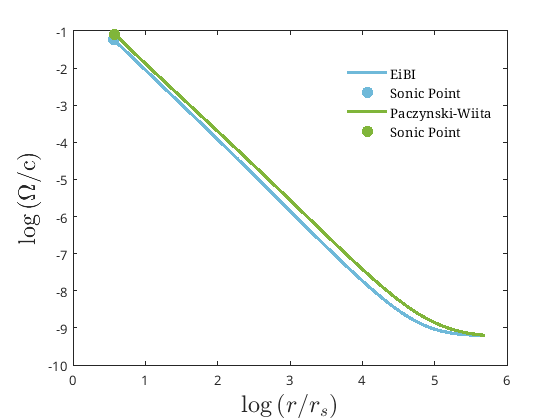}
    \hfill
    \includegraphics[width=0.32\linewidth]{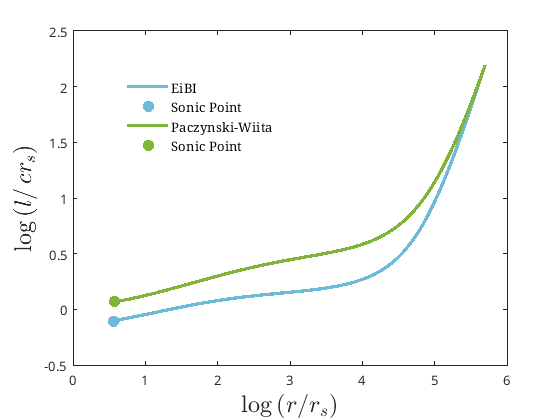}

    \vspace{0.1cm}

    \includegraphics[width=0.31\linewidth]{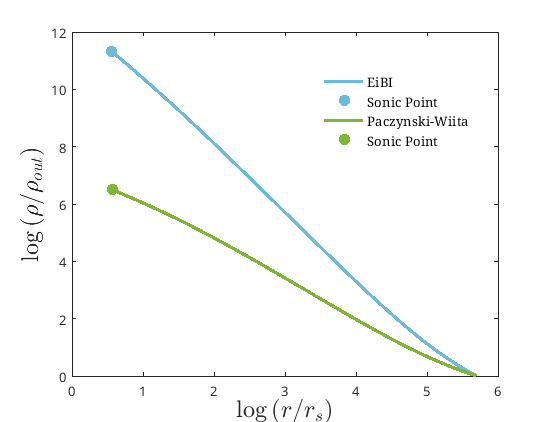}
    \hfill
    \includegraphics[width=0.32\linewidth]{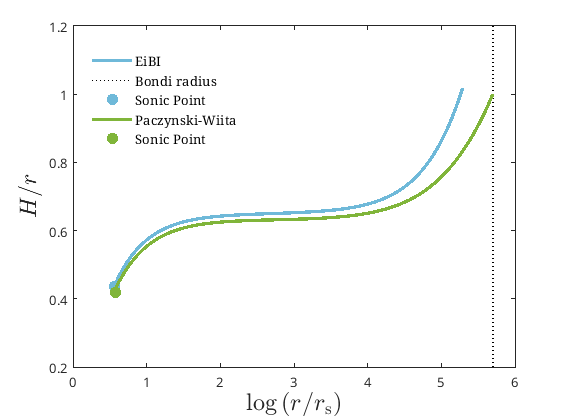}
    \hfill
    \includegraphics[width=0.32\linewidth]{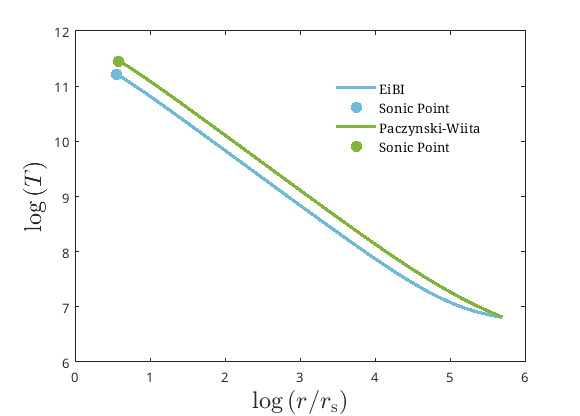}

    \caption{
    Radial profiles of the radial velocity $v$, sound speed $c_s$, angular velocity $\Omega$, specific angular momentum $l$, density $\rho$, disk thickness $H$, and temperature for \ac{ADAF}-like accretion flows in the \ac{PW} (black) and Palatini (red) gravitational potentials. Filled circles denote the sonic points. 
    Sonic point is $r_{sp} = 3.7673 \, r_s$ for standard \ac{ADAF}-like, and is $r_{sp} = 3.6072 \, r_s$. For Palatini gravity $s = -1$, $r_c = 10^{-32} \, r_s$;
$r_q = 10^{-30} \, r_s$. These values are motivated by the constraints on the electric charge in \cite{Zajacek_2019}.}
    \label{fig:six_plots}
\end{figure*}

\begin{figure}[htbp]
    \centering
    \includegraphics[width=0.7\linewidth]{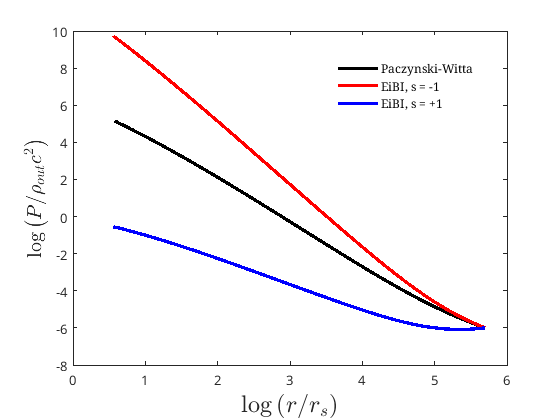} \\
    \includegraphics[width=0.7\linewidth]{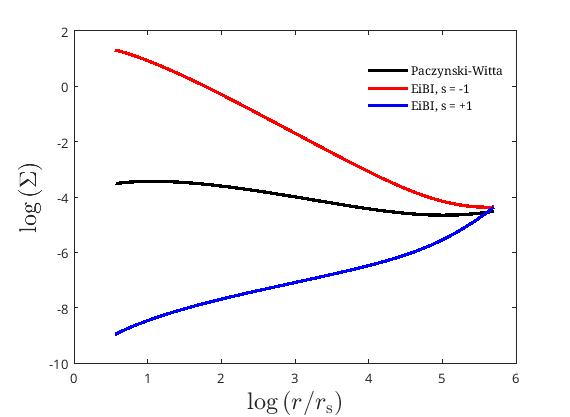}

    \caption{Comparison of the radial profiles of the density (top) and surface density (bottom) for the standard \ac{PW} potential (black) and the Palatini gravity model with $s=-1$ (red) and $s=+1$ (blue). The solutions extend from the Bondi radius ($ \sim 10^{5} \, r_s$) to the sonic point. Palatini gravity significantly modifies the gas distribution, leading to enhanced density for $s=-1$ and reduced density for $s=+1$ compared with the standard potential.}
    \label{s__plots}
\end{figure}

\subsection{Method}

Combining the continuity, radial momentum, angular momentum, and energy equations, given by Eqs. \eqref{Eq: Cont eq}, \eqref{eq: radial equation}, \eqref{int: angualr eq}, and \eqref{energy equation}, respectively, yields a system of three coupled first-order ordinary differential equations for the radial velocity, $v_r(r)$, the angular velocity, $\Omega(r)$, and the sound speed, $c_s(r)$. This system is singular at the sonic point, where the flow transitions from subsonic to supersonic. 


\begin{equation}
    \frac{d ln |v_r|}{dr} = \frac{\mathcal{N}}{\mathcal{D}} \ .
\end{equation}

The solution is regular at the sonic point when the numerator and denominator of $dv_r/dr$ appearing in the previous equation vanish simultaneously, that is

\begin{eqnarray}
    \mathcal{N} &=& (\Omega_K^2 -\Omega^2) - \frac{2\gamma c_s^2}{r} + \frac{(\gamma-1)(\Omega r^2 - j)^2v_r}{\alpha r^3 c_s} = 0 \ , \nonumber \\
    \mathcal{D} &=& (\gamma c_s^2 - v_r^2)  = 0\ . 
\end{eqnarray}

This condition, together with the requirement that the flow smoothly matches quasi-Keplerian rotation at large radii, fixes two eigenvalues of the problem: the location of the sonic point, $r_{\rm sp}$, and the specific angular momentum swallowed by the \ac{BH}, $j$.

We impose boundary conditions at the outer edge of the flow, near the Bondi radius ($r_{\rm out}\sim10^{5}\,r_s$), following \citet{Ranjbar2022}, and integrate inward to the sonic point $r_{\rm sp}$. The sound speed and density at the outer boundary are held constant. We adopt $c_{\rm out}=10^{-3}\,c$, which corresponds to a temperature of $T\simeq6.5\times10^{6}\,\mathrm{K}$ for a mean molecular weight $\mu=0.6$, using $c_s=\sqrt{\gamma k_B T/\mu m_p}$. This is a reasonable value for the interstellar medium at the center of a galaxy \citep{Narayan2011}.

We solve this two-point boundary value/eigenvalue problem using a relaxation method \citep{Press1992}, following its recent application to similar low-angular-momentum \ac{ADAF}-like flows 
\citep{Narayan2011, Ranjbar2022, Ranjbar2025}. Starting from an initial guess for the solution and for the eigenvalues $r_{\rm sp}$ and $j$, the solution is iteratively corrected. Five boundary conditions are required, but only the outer boundary $r_{\rm out}$ can be specified directly; the critical radius $r_{\rm sp}$ is an eigenvalue that must be determined as part of the solution, which makes this a free boundary problem. We reduce it to a standard boundary value problem by introducing a new dependent variable $y_{4}$ with $dy_{4}/dr = 0$ and a new independent variable $t$, defined by $\ln(r) - \ln(r_{\rm c}) = t y_{4}$, with $0 \leq t \leq 1$. Both boundaries are then fixed at $t = 0$ and $t = 1$, so the system can be solved with the standard relaxation method.

\section{Results}\label{sec:4}

In this section, we present the global transonic solutions obtained using the EiBI gravitational potential and compare them with the standard \ac{PW} potential, which are depicted in Fig. \ref{fig:PW}. The results of the solutions are shown in Figure~\ref{fig:six_plots} the radial profiles of the radial velocity, sound speed, specific angular momentum, density, disk thickness, and temperature. The solutions extend from the outer boundary at the Bondi radius ($10^{5}\,r_s$) inward to the sonic point at $r_{\rm sp}\simeq3\,r_s$, marked by filled circles for each solution. The figure is plotted for $s=-1$. This choice was motivated by the original paper, since the resulting geometry is regular, namely geodesically complete. Adopting $s=1$ leaves all of these quantities unchanged, with the curves overlapping those for $s=-1$, except for the density, which differs between the two cases.

Compared with the standard \ac{PW} case, the EiBI gravity solution predicts lower $v_r$, $c_s$, $l$, and $T$ over most of the accretion flow. In contrast, $\rho$ and $H$ increase systematically. These differences indicate that the modified gravitational potential slows the inward motion of the gas, allowing matter to accumulate more efficiently and producing a denser and geometrically thicker accretion flow. The reduction in the infall velocity also decreases compressional heating, resulting in a lower gas temperature. Although the overall behaviour of the solution remains similar to the \ac{ADAF}-like transonic flow, EiBI gravity modifies the quantitative structure of the accretion flow, particularly in the inner regions close to the \ac{BH}, as expected.

\begin{figure*}[htbp]
    \centering
    \includegraphics[width=0.6\linewidth]{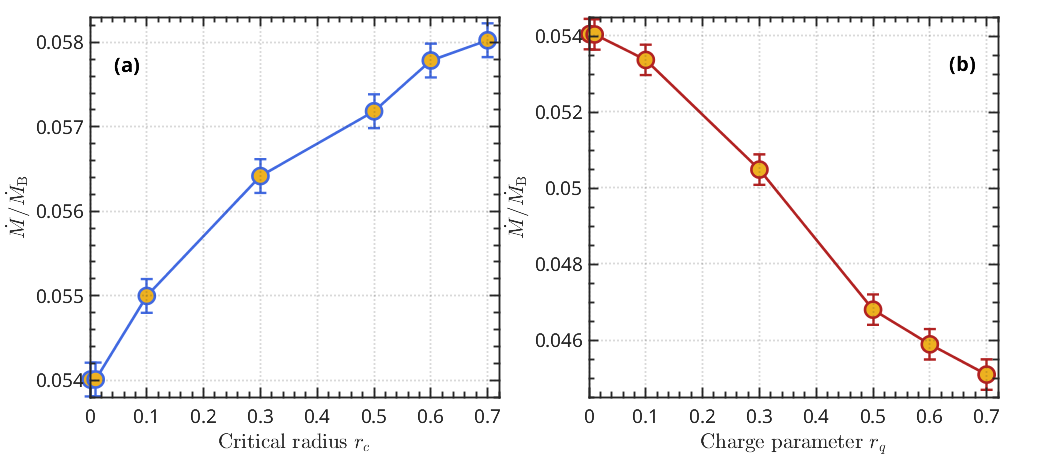} 

    \caption{
Dependence of the normalised mass accretion rate on the EiBI gravity parameters. Panel (a) shows the variation of $\dot{M}/\dot{M}_{\rm B}$ with the critical radius $r_c$, while panel (b) presents the dependence on the charge parameter $r_q$. Error bars represent the numerical uncertainties of the transonic solutions. Increasing $r_c$ enhances the mass accretion rate, whereas increasing $r_q$ suppresses it.
}
    \label{mdot_plots}
\end{figure*}

Figure \ref{s__plots} compares the radial distributions of the density and surface density obtained using the standard \ac{PW} potential and the EiBI gravity model for the only two possible values of the parameter $s$. The solutions extend from the Bondi radius ($10^{5} \, r_s$) to the transonic point.

The modified gravitational potential has a significant impact on the gas distribution. For the case $s=-1$, both the density and the surface density are systematically higher than those of the standard \ac{PW} solution throughout most of the accretion flow. This behaviour is a direct consequence of the reduced radial inflow velocity, which allows gas to accumulate more efficiently in the disk. In contrast, the case $s=+1$ produces lower densities, indicating a less massive accretion flow.

These results demonstrate that the EiBI gravitational potential can substantially alter the structure of \ac{ADAF}-like accretion flows while preserving the solution's overall transonic nature. The dependence on the parameter $s$ is particularly pronounced, suggesting that different branches of the EiBI solution lead to distinct accretion flow properties.
To investigate the influence of the EiBI gravity parameters on the accretion flow, Figure \ref{mdot_plots} presents the variation of the normalised mass accretion rate, $\dot{M}/\dot{M}_{\rm B}$, as a function of the critical radius $r_c$ and the charge parameter $r_q$. The values of $r_c$ and $r_q$ adopted in Figure~\ref{mdot_plots} are illustrative rather than physically realistic. They are chosen solely to demonstrate the dependence of the mass accretion rate on these parameters.

As shown in panel (a) of Fig. \ref{mdot_plots}, the normalised accretion rate increases monotonically with increasing $r_c$. A larger core radius weakens the modifications to the gravitational potential near the compact object, allowing a slightly larger fraction of the Bondi inflow to be accreted. In contrast, panel (b) shows that increasing the charge parameter $r_q$ systematically decreases the mass accretion rate. The stronger modification of the gravitational potential slows the radial inflow, thereby reducing the amount of matter reaching the central object.\\
Although the variations are modest, they demonstrate that the accretion rate is sensitive to the EiBI gravity parameters. This dependence provides a possible observational signature through measurements of mass accretion rates in low-luminosity accreting \acp{BH}.
The normalised mass accretion rate increases from approximately
$\dot{M}/\dot{M}_{\rm B}\simeq0.054$ to $\simeq0.058$ as the core radius increases over the explored range. Conversely, increasing the charge parameter reduces the accretion rate from $\simeq0.054$ to $\simeq0.045$. These results indicate that the charge parameter has a stronger influence on the accretion dynamics than the core radius within the considered parameter space.\\

\section{discussion}\label{sec:5}

In this paper, we investigate the effect of Palatini Eddington-inspired Born-Infeld (EiBI) gravity on the structure of a transonic, low-angular-momentum \ac{ADAF}-like accretion flow around a static, spherically symmetric \ac{BH}. Our solutions describe the subsonic region of the flow, extending from the outer boundary near the Bondi radius ($10^5\,r_s$) to the sonic point.

Earlier studies of this class of accretion flow have relied either on the Newtonian potential or on the \ac{PW} pseudo-Newtonian potential, which reproduces several relativistic features of the Schwarzschild spacetime without solving the full field equations. In this work, we instead derive the pseudo-Newtonian potential directly from the geometry of an exact EiBI solution, extending the general-relativistic content captured by the standard \ac{PW} approach to a modified-gravity background.

We find that, relative to the standard \ac{PW} potential, the EiBI potential for $s=-1$ systematically lowers the radial velocity, sound speed, specific angular momentum, and gas temperature throughout most of the flow, while increasing the density and disk thickness. The sonic point itself shifts inward, from $r_{\rm sp} \simeq 3.7673\,r_s$ in the standard case to $r_{\rm sp}\simeq 3.6072\,r_s$ under EiBI gravity. Physically, the modified potential weakens the effective inward pull on the gas near the compact object, slowing the infall, reducing compressional heating, and allowing matter to accumulate into a denser, geometrically thicker disk. This trend reverses for the $s=+1$ branch, which instead produces a less massive, lower-density flow. This indicates that the sign of $s$, which distinguishes the \ac{BH} and wormhole-like branches of the EiBI solution, has a qualitative rather than merely quantitative effect on the accretion structure. This reversal reflects the different near-field geometry of the two branches: the $s = +1$ solution describes a singular black hole structure in which the potential does not develop the same enhancement close to the centre as in the $s = -1$ case, corresponding to the wormhole case, the gas experiences comparatively less resistance to infall, resulting in a more efficient, lower-density accretion flow.

The mass accretion rate is likewise sensitive to the two EiBI parameters. Increasing the core radius $r_c$ raises $\dot{M}/\dot{M}_{\rm B}$ (from $\simeq0.054$ to $\simeq0.058$) over the explored range, since a larger $r_c$ weakens the deviation from the standard potential near the compact object. Increasing the charge-like parameter $r_q$, by contrast, suppresses $\dot{M}/\dot{M}_{\rm B}$ (from $\simeq0.054$ to $\simeq0.045$), reflecting a stronger modification of the potential that slows the radial inflow. The larger sensitivity to $r_q$ than to $r_c$ suggests that observed mass accretion rates in low-luminosity accreting \acp{BH} could, in principle, be used to constrain this parameter more tightly than the core radius, offering a possible observational handle on departures from standard gravity in these systems.

We emphasise that the present results are restricted to a non-rotating, spherically symmetric background, low angular momentum, and the specific EiBI parameter ranges explored here, which were chosen for illustrative rather than observationally motivated purposes. A more astrophysically realistic exploration of the parameter space, together with a comparison to observed accretion rates in systems such as M87* and Sgr A*, is left for future work.

As a natural extension of this work, we plan to generalise the calculation to a rotating (Kerr-like) \ac{BH} within the same EiBI framework, using a PW-type potential adapted for rotating spacetimes. This would allow us to assess how the combination of \ac{BH} spin and EiBI modifications jointly affects the structure of low-angular-momentum accretion flows. Additionally, we want to include an analysis of the accretion disk's emission intensity profile, as well as the region very close to the horizon, which was removed in this work.

\section*{Acknowledgments}

M.G. acknowledges support by CIDMA under the FCT Multi-Annual Financing Program for R\&D units, the Hanse-Wissenschaftskolleg (HWK) - Institute for Advanced Study (Delmenhorst, Germany) through a Junior Fellowship and the Estonian Research Council grant MOB3JD1241.

\bibliography{ref}

\end{document}